\documentclass[12pt,preprint]{aastex}
\usepackage{amssymb}
\newcommand{\myemail}{hayato@crab.riken.jp}
\shortauthors{Hayato et al.}

\begin{document}

\title{Discovery of a Compact X-ray Source \\
	in the LMC Supernova Remnant N23 with {\it Chandra}} 
\author{Asami Hayato\altaffilmark{1,2}, Aya Bamba\altaffilmark{2}, 
	Toru Tamagawa\altaffilmark{2,1}, and Kiyoshi Kawabata\altaffilmark{1}}
\altaffiltext{1}{Department of Physics, Faculty of Science, 
	Tokyo University of Science, 1-3, Kagurazaka, Shinjyuku-ku, 
	Tokyo 162-8601, Japan; \myemail}
\altaffiltext{2}{RIKEN (Institute of Physical and Chemical Research), 2-1, 
	Hirosawa, Wako, Saitama, 351-0198, Japan}

\begin{abstract}

An X-ray compact source was discovered with {\it Chandra} in a
supernova remnant (SNR) N23, located in the Large Magellanic
Cloud. The compact source (CXOU J050552.3$-$680141) is seen in only
the hard band ($>$ 2 keV) image of N23, while the soft band image ($<$
2 keV) shows diffuse emission of the SNR, with an extent of
$\sim 60 \arcsec \times 80\arcsec$. The compact source is located at
almost the center of N23, and there is no identifiable object for the
source from previous observations at any other wavelength. The source
spectrum is best explained by a power-law model with a photon index of
$\Gamma$ = 2.2 $^{+0.5}_{-0.3}$ and an absorption-corrected luminosity
of $L_{x}$ = 1.0 $\times$ 10$^{34}$ ergs s$^{-1}$ in the 0.5--10 keV
band for a distance of 50 kpc. Neither pulsation nor time variability
of the source was detected with this observation with a time
resolution of 3.2 sec. These results correspond with those of
\citet{hughes06}, who carried out analysis independently around the
same time as our work.

Based on information from the best-fit power-law model, we suggest
that the source emission is most likely from a rotation-powered pulsar
and/or a pulsar wind nebula. It is generally inferred that the
progenitor of N23 is a core-collapsed massive star.

\end{abstract}

\section{Introduction}

The Large Magellanic Cloud (LMC) quite well meet the requirements of
a systematic study of supernova remnants (SNRs), in view of the facts
concerning the known distance (50 kpc; Feast 1999) and the small
absorption because it is the nearest face-on galaxy to us. Over 30
SNRs have been discovered in the LMC so far, and categorized
\citep{williams99,hughes98}. Besides, due to the limited spatial 
resolution of previous observations, it is reported that only six of
pulsar/SNR associations in the LMC, PSR B0540$-$69 in N158A
\citep{gotthelf00}, PSR J0537$-$6910 in N157B \citep{marshall98}, a
pulsar in N206 \citep{klinger02}, SGR 0526$-$66 in N49
\citep{kulkarni03,rothschild94}, XMMU J053559.3$-$673509 in DEM L241
\citep{bamba06}, and a pulsar wind nebula (PWN) in SNR B0453$-$68.5
\citep{gaensler03a}. A PWN is a synchrotron emission from
high pressure plasma swept by the pulsar's relativistic wind.  Even
though all of PWNe are powered by a pulsar, there is a case of a PWN
without detection of a pulsar \citep{gaensler03a}. The pulsar/SNR or
PWN/SNR associations provide meaningful information, such as the age
of the systems, and the type of remnants.
 
The SNR N23 (0506$-$68.0) is one of the SNRs in the LMC that
\citet{milne80} identified as an SNR with radio observations at 5 and
14.7 GHz for the first time. It is classified by its morphology as
being a middle-aged ($\sim$3000 year) and an irregular-shaped SNR in
\citet{hughes98}. SNR N23 has not been well studied, so
there is still no absolute solution for its explosion type (a
core-collapse or a Ia). \citet{cohen88} insisted the possibility
that N23 is a remnant from a massive star SN, since a CO
cloud with high-density OB stars was observed within $\sim$30 pc of
N23. On the other hand, \citet{banas97} suggested that the remnant
does not have any association with others according to their own
observations with better spatial resolutions than that of the Cohen
survey.

We present the results of an X-ray observation of N23 with {\it
Chandra} in this paper. The observation details are summarized in
$\S$2, and the analysis results in $\S$3. Using those results, we
discuss the source discovered at the center of the remnant in $\S$4,
and $\S$5 is a summary of this paper. \citet{hughes06} reported the
same results based on their independent analysis.

\section{Observation Details}

An observation of SNR N23 was performed with the {\it Chandra}
\citep{weisskopf02} ACIS-S detector \citep{garmire00} on December 29, 2002
(observation ID = 2762). The operation was in a timed exposure very
faint mode, and it provided a time resolution of 3.24 sec. Data
reduction was done using {\it Chandra} Interactive Analysis of
Observation (CIAO) version 3.1 and CALDB version 2.28. From the Level
1 processed events that {\it Chandra} X-Ray Center gives, {\it
Advanced Satellite for Cosmology and Astrophysics} ({\it ASCA}) grades
of 0, 2, 3, 4, and 6 were selected. After that, the data was filtered
for good time intervals and, consequently, the total usable exposure
time was 37 ksec.
\section{Results}

\subsection{Image Analysis} 
\label{sec:imgana}

Figure~\ref{fig:img} shows ACIS X-ray images in the (a) 0.5--2.0 keV
and (b) 2.0--8.0 keV bands. It demonstrates that the diffuse emission
extend $\sim 60 \arcsec \times 80 \arcsec$ (15 pc $\times$ 19 pc for a
distance of 50 kpc) in the soft X-ray band. The eastern part of the
SNR is quite bright, whereas the northern and western parts are
not. The lack of uniformity of the morphology might be due to a
molecular cloud sitting southeast of N23 \citep{banas97,math85}.

In the hard X-ray image, a compact source (CXOU J050552.3$-$680141) is
found in the SNR, although diffuse emission is not. It was detected by
the {\tt wavdetect} command in CIAO with a significance of 11 $\sigma$
in the 2.0$-$8.0 keV band. The contour of 3 $\sigma$ detection has an
elliptical shape (1\farcs5 $\times$ 1\farcs3), and is slightly larger
than the point spread function size at the source position of ACIS;
however, it cannot be concluded that the source significantly extends,
since there are non-uniform and complicated diffuse emissions over the
source. Its location was identified as R.A.= 05$^{\rm h}05^{\rm
m}$52\fs3, decl.=$-$68\arcdeg01\arcmin41\farcs2 (J2000) with an
uncertainty of less than a pixel size (0.5 arcsec). No counterpart
object in this position has been found with the SIMBAD data base; in
addition, there is no report about the compact source in either
optical \citep{math85} or radio \citep{dickel98,banas97} band
observations. An optical and a radio limits are magnitude $>$ 17 and
$<$ $\sim$2mJy respectively \citep{hughes06,dickel98}. We concentrate
our discussion on this compact object in this paper, and call it
``Source'' hereinafter.

\subsection{Spectral Analysis}
\label{sec:specana}

\begin{figure}
\plottwo{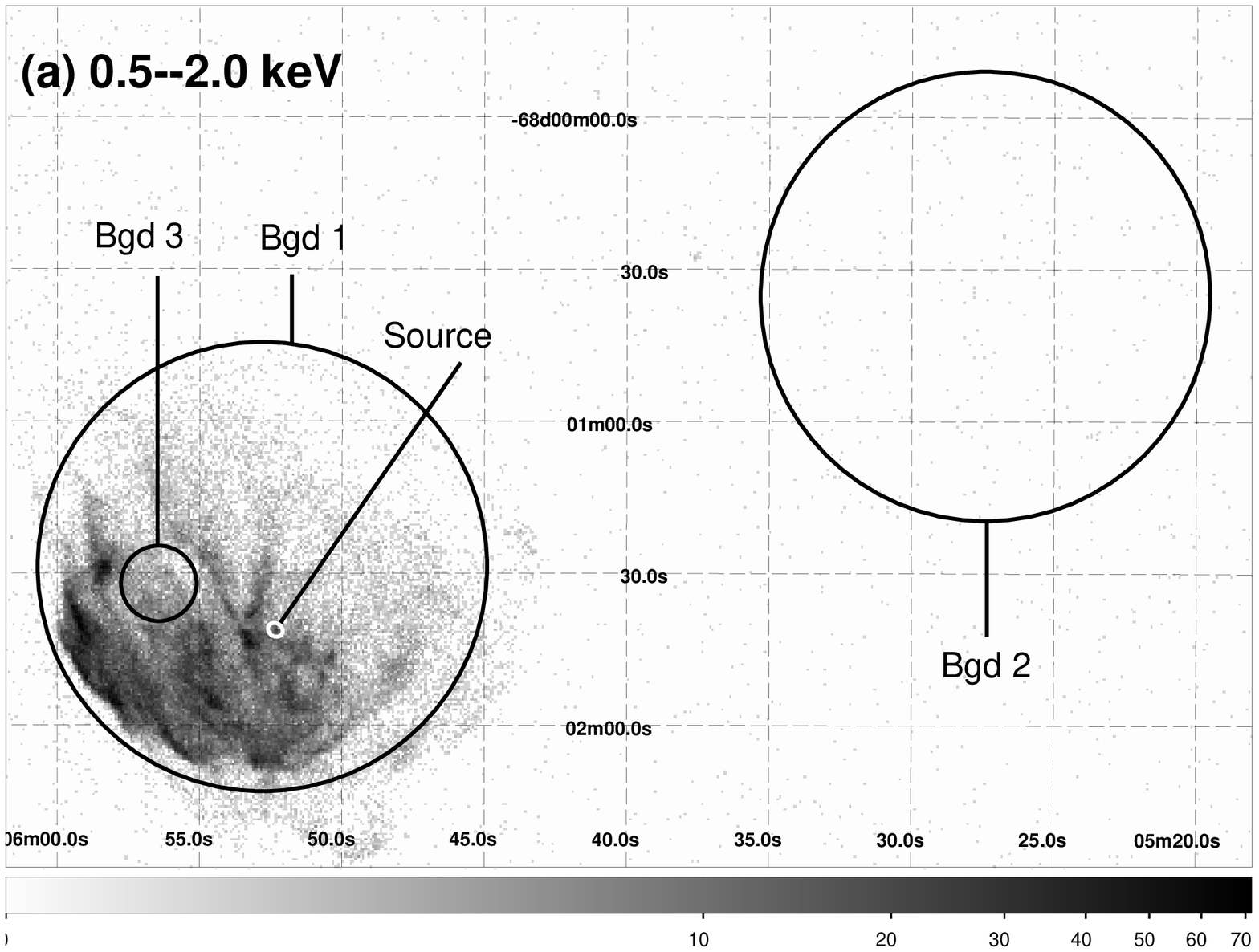}{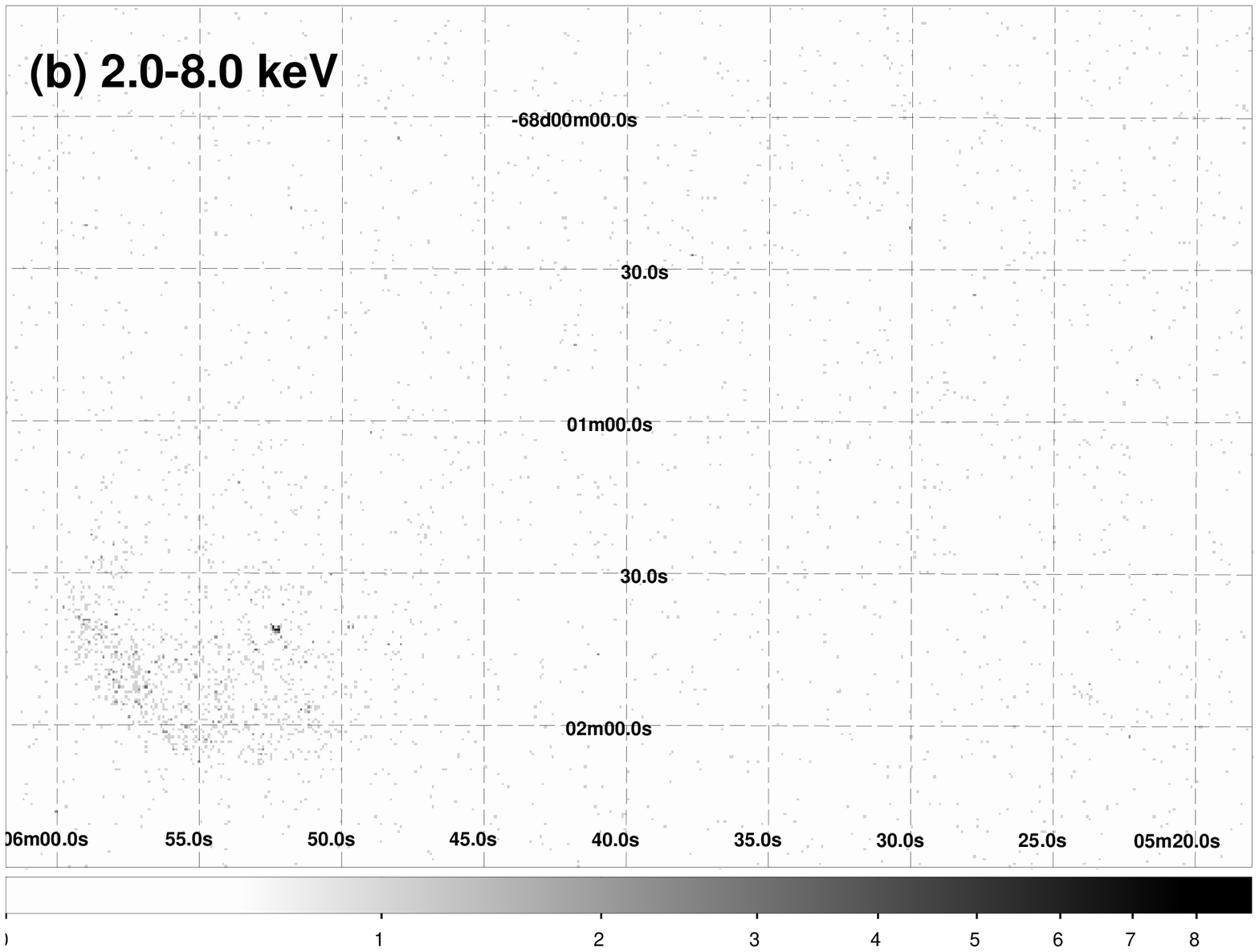}
\caption{{\it Chandra} ACIS X-ray images of SNR N23 in the 
(a) 0.5--2.0 keV and (b) 2.0--8.0 keV bands in a logarithmic scale. 
The source and background regions are shown in the panel (a). 
The coordinates are in J2000. \label{fig:img}}
\end{figure}

We obtained the spectrum of Source with photons in the elliptical
region at Source, described in $\S$\ref{sec:imgana} (Figure
\ref{fig:img} (a)). There were 285 and 37 events in the region in the
0.5--2.0 keV and 2.0--8.0 keV bands, respectively, and the corresponding
count rates were 4.4 and 0.57 counts sec$^{-1}$ arcmin$^{-2}$ with an
exposure time of 37 ksec.

Because the background of Source surely consists of the thermal emission of
SNR itself, we tested three backgrounds: the first was a circle
that included the whole SNR (Bgd 1), the second was a circle from an
extraneous region (Bgd 2), and the third was a circle that had 
almost the same surface brightness as the surrounding region of Source
(Bgd 3), as shown in Figure \ref{fig:img} (a). 
 
\begin{figure}
\plotone{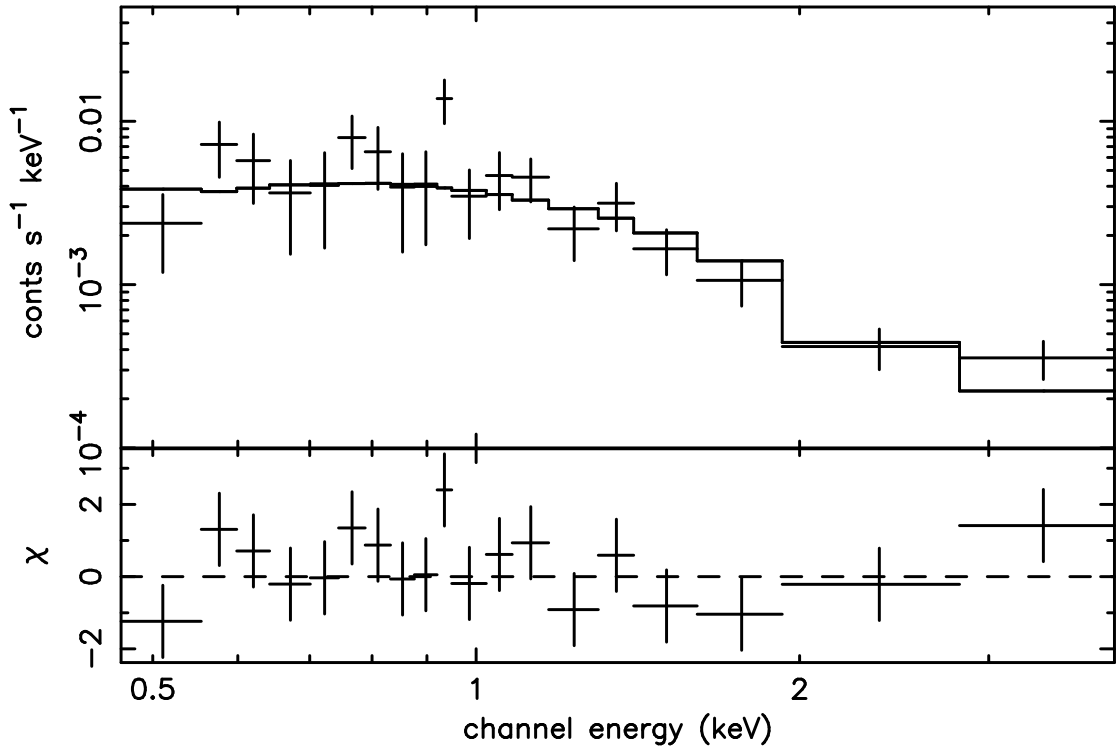}
\caption{The background (Bgd 1) subtracted spectrum of the source (crosses) 
with the best fit model, the power-law function (solid line). 
Residuals are shown in the bottom panel.
\label{fig:spec}}
\end{figure}

We fitted three spectra with an absorbed power-law function, blackbody
model, and plasma (Mekal) model
\citep{mewe85,mewe86,kaastra92,liedahl95}, assuming that photons are
absorbed in Galaxy ($N{_{\rm H}}^{\rm gal}$) and the LMC ($N{_{\rm
H}}^{\rm LMC}$). The calculation of absorption columns was done using
cross sections obtained by \citet{morrison83} for $N{_{\rm H}}^{\rm gal}$
and \citet{baluc92} for $N{_{\rm H}}^{\rm LMC}$, under the assumption
that the abundances are 1 solar for $N{_{\rm H}}^{\rm gal}$
\citep{anders89} and 0.3 solar for $N{_{\rm H}}^{\rm LMC}$
\citep{russ92,hughes98}. We froze $N{_{\rm H}}^{\rm gal}$ to 5.6 $\times
10^{20} {\rm cm}^{-2}$ \citep{dickey90}, while we left $N{_{\rm
H}}^{\rm LMC}$ a free parameter. Table \ref{tbl:fit_results} gives the
results of the fittings. Even though we tested three models with three
different background regions, it did not matter for the results of the
absorption-corrected flux, $\sim10^{-14}$ ergs cm$^{-2}$ s
$^{-1}$. However, only the power-law fitting of the spectrum with the
Bgd 1 gave an acceptable reduced $\chi^2$. Figure \ref{fig:spec}
represents the best-fit power-law model and data with Bgd 1.

\subsection{Timing Analysis}

\begin{figure}
\plotone{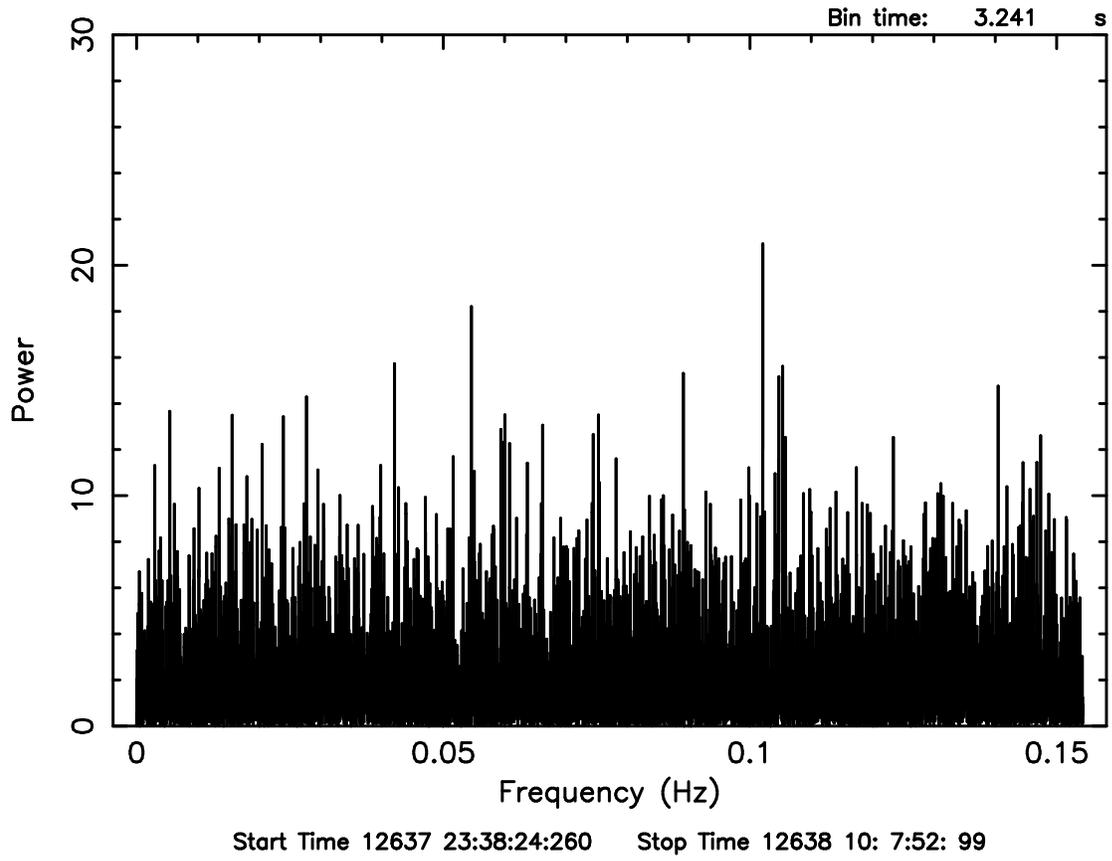}
\caption{The power spectrum of Source in the 2.0--8.0 keV band.
\label{fig:powspec}}
\end{figure}

Coherent pulsations of Source were searched. Selected photons are from
the region of Source (see $\S$\ref{sec:specana}) and in the 2.0--8.0
keV band. The arrival times of photons were corrected with the
{\tt axbary} command in CIAO. Figure~\ref{fig:powspec} gives a
power-density spectrum of Source made based on a fast Fourier
transform (FFT) algorithm.  Even the highest peak dose not protrude
significantly from the white noise; thus, we consider that no
significant pulsation was detected. Note that the time resolution of
ACIS is only 3.24 sec in this observation.

\begin{figure}
\plotone{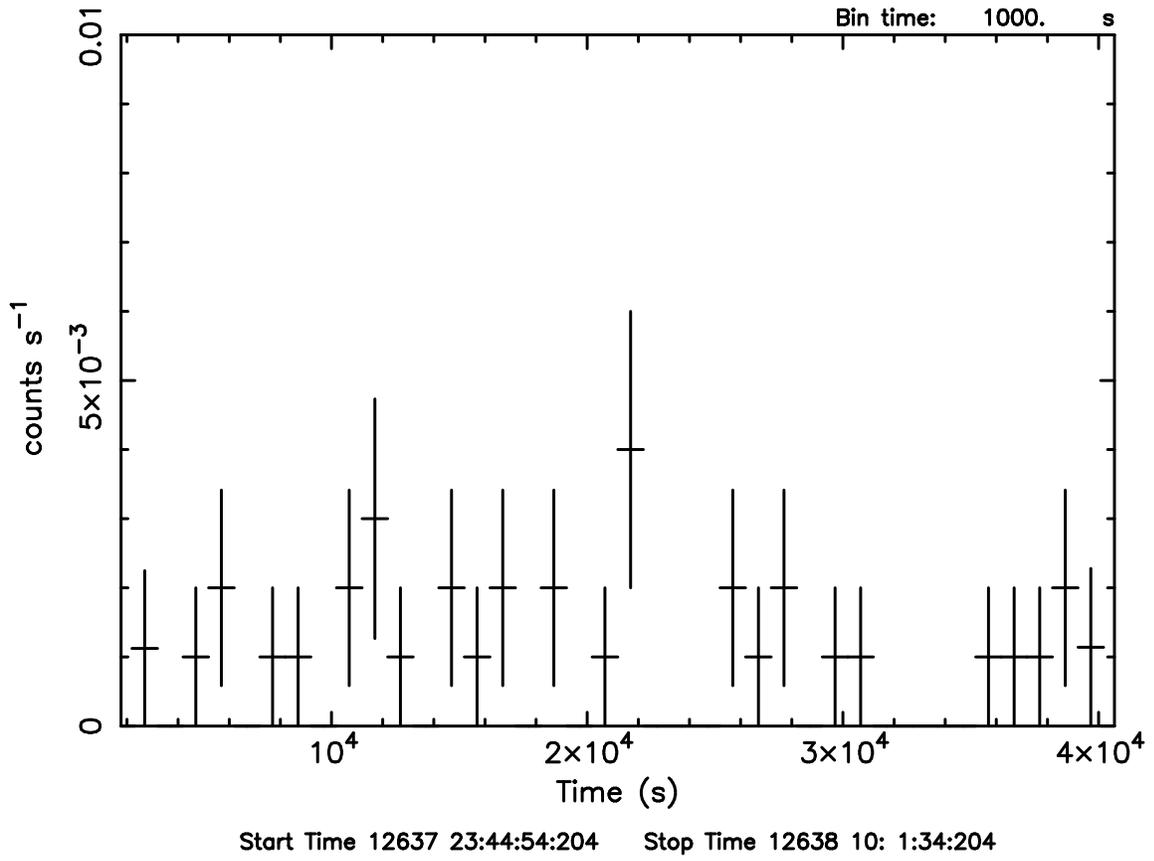}
\caption{The X-ray light curve of Source in the 2.0--8.0 keV band, binned with 1000 s. 
The background is not subtracted.\label{fig:lc}}
\end{figure}

We also searched for the time variability of Source. We made a light
curve of the entire observation time, as demonstrated in
Figure~\ref{fig:lc}. A statistical analysis with the Kolmogorov-Smirnov
test shows 53\% for the probability of its constancy. As a result, we
have concluded that Source has no significant time variability.

\section{Discussion}

Although a physical association between Source and the SNR N23 is
strongly supported by the fact that the Source location is almost at
the center of the SNR, our fitting results of the $N{_{\rm H}}^{\rm
LMC}$ do not determine whether Source is in LMC or not due to the
error. We thus verified the case that Source is (1) an object in
Galaxy, (2) the object in LMC is unrelated with SNR, (3) a background
active galactic nucleus (AGN), (4) the stellar remnant of N23.

\subsection{If Source is in Galaxy}

If Source is in Galaxy, possible candidates are an active RS CVn
binary ($L_x$=29$-$32), a Cataclysmic Variables (CV; $L_x$=30$-$33), a
Young Stellar Object (YSO; $L_x$=29$-$31), or a Low Mass X-ray Binary
(LMXB; $L_x=$30$-$39), because their luminosities are consistent with the
Source luminosity, assuming that Source is in Galaxy.
 
Since plasma models with any of three backgrounds did not provide an
acceptable reduced $\chi^2$, we concluded that Source is unlikely to
be a RS CVn, a CV, or a YSO, which are well explained by plasma
models. The emission from a LMXB can explained by a power-law model,
but the photon index is typically 1--2 \citep{muno03}, which is harder
than the best-fit power-law model of Source.

As a result, we can say that Source would not be a 
considerable object in the case that Source is in Galaxy.

\subsection{Source possibility of being a HMXB in the LMC}

If Source is in the LMC and unrelated with the SNR, it could be a
HMXB.  The luminosity of HMXB (log$ L_x$ = 33--38; Muno et al. 2003)
corresponds to the luminosity of Source, if Source is in the LMC. The
observed spectrum of a HMXB is explained with a photon index of
0.5--2.5 \citep{muno03}, which is also the same as the best-fit
power-law of Source. However, the HMXB should be observed with the
companion star in the optical band, and the expected optical limit of
the companion star in the LMC is magnitude 16. The work by
\citet{hughes06} using the data of the MACHO Project indicates that the
optical limit in the Source region is $>$ 17. We can thus reject the
possibility of being a HMXB in LMC.

\subsection{Source possibility of being an AGN}

The spectrum of an AGN has a photon index of $\sim$1--3
\citep{beckmann06}, and the photon index of Source is in between. 
We calculated the expected number of AGNs in the back of N23 using a
log{\it N}$-$log{\it S} function measured by the {\it ROSAT} deep
survey in the Lockman Hole \citep{hasinger98}. As a result, 24 AGNs,
the flux of which is equal or higher than Source, are spread per
square degree in the sky. Then, because the possibility of the AGN
coincidence in the SNR region (elliptical; $\sim 60\arcsec \times
80\arcsec$) is only 0.8\%, we can hardly find an AGN in the back of
N23.

\subsection{Source as a stellar remnant of SNR N23}

It is highly possible that Source associates with the SNR due to its
location. The existence of an associating stellar remnant suggests
that N23 is outcome of gravitational core-collapse of a massive star.

The possible candidates for Source as a stellar remnant of N23 would
be a neutron star, a compact central object (CCO), an isolated pulsar,
or a PWN. 

The emission by a neutron star can be explained by a blackbody model
with $kT$ = 0.15--0.25 keV \citep{page96,page98}, but the blackbody
model was not able to represent the Source radiation with any
backgrounds. Thus, Source is hardly said to be a usual neutron star.

\begin{figure}
\begin{center}
\plotone{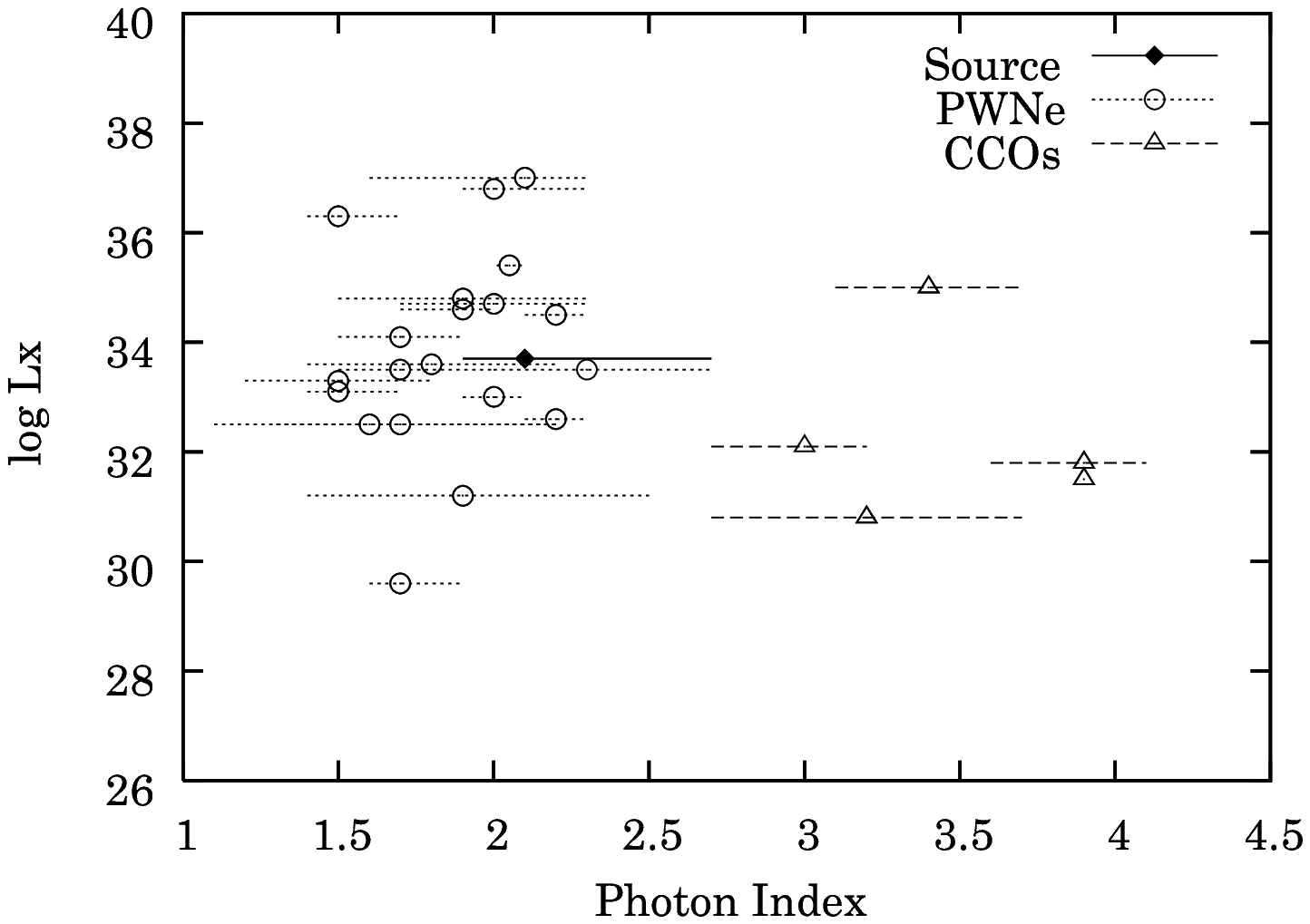}
\caption{Relationships of photon indices and luminosities of Source 
(a solid line with a diamond), PWNe (doted lines with circles), and
CCOs (dashed lines with triangles).  Referred PWNe are in SNR
G18.0-0.7 (assumed distance D = 4 kpc; \citet{gaensler03b}), 
in N157B (D = 50 kpc; \citet{marshall98,wang98}), 
PSR B0540$-$69 (D = 50 kpc; \citet{kaaret01}),
in G11.2-0.3 (D = 5.0 kpc; \citet{roberts03}), 
Crab (D = 2.0 kpc; \citet{petre02}), 
Vela (D = 0.3 kpc; \citet{pavlov01}), 
Geminga (D = 0.16 kpc; \citet{caraveo03}), 
PSR J2229$+$6114 (D = 3.0 kpc; \citet{halpern01}), 
PSR 1105$-$6107 (D = 7 kpc; \citet{gotthelf98}),
PSR 1706$-$44 (D = 2.5 kpc; \citet{finley98}), 
PSR B1757$-$24 (D = 5 kpc; \citet{kaspi01}), 
in 3C58 (D = 2.6 kpc; \citet{murray02,torii00}), 
PSR B1957$+$20 (D = 1.5 kpc; \citet{stappers03}), 
PSR J2021$+$3651 (D = 10 kpc; \citet{hessels04}), 
PSR J1747$-$2958 (D = 5 kpc; \citet{gaensler04}),
in SNR G292.0$+$1.8 (D = 4.8 kpc; \citet{hughes01}), 
in W44 (D = 2.02 kpc; \citet{petre02}), 
PSR J1930$+$1852 (D = 5 kpc; \citet{lu02}), 
PSR B0458$-$685 (D = 50 kpc; \citet{gaensler04}),
and PSR B1509$-$58 (D = 5.2 kpc; \citet{gaensler02}). 
The PWNe list of \citet{cheng04} are used. 
CCOs are in Cas A (D = 3.4 kpc; \citet{mereghetti02}), 
in G226.2$-$1.2 (D = 0.4 kpc; \citet{pavlov01}), 
in G296.5$+$10.0 (D = 1.5 kpc; \citet{mereghetti96}), 
in RCW 103 (D = 3.3 kpc; \citet{gotthelf99}), 
and in Kes 73 (D = 7.0 kpc; \citet{gotthelf97}). 
Error bars indicate 90\% confidence.
\label{fig:pi_lum}}
\end{center}
\end{figure}

CCOs have recently been discovered in several SNRs, like the central
source in Cassiopeia A \citep{chakrabarty01} or in Puppis A
\citep{zavlin99}. CCOs' spectra can be fitted with either a blackbody
or a power-law model, but the blackbody temperatures of CCOs are
excessively high ($kT$ = $\sim$0.3--0.6 keV; Chakrabarty et al. 2001,
Kaspi \& Roberts 2004) to consider that they are the usual neutron
stars; also, the photon indices of power-law fitting is too soft
($\Gamma$ = $\sim$2.6--4.1; Kaspi \& Roberts 2004) to be pulsars. The
luminosity of CCOs is log{\it L}$_x$ = 31--35 ergs s$^{-1}$, as shown
by the triangle symbols in Figure~\ref{fig:pi_lum}. Even though the
Source luminosity is consistent with those of CCOs, Source could not
be acceptably fitted with a blackbody model, and the photon index of a
power-law model is harder than the typical photon index of CCOs. Thus,
we consider that Source in N23 seems not to be a CCO.

A pulsar typically has the power-law spectrum with a photon index of
$\Gamma$ = $\sim$0.6--1.9 \citep{gotthelf00}, which is harder than the
best-fit photon index of Source. Meanwhile, for a PWN, we would
observe a softer photon index, generically, $\Gamma$ = $\sim$1.1--2.3
\citep{cheng04}. The luminosity of the PWN is log$ L_x$ = 29--37,
which is brighter than that of only the pulsar \citep{cheng04}. This
is indicated in Figure~\ref{fig:pi_lum} by circle symbols. Due to the
spectrum agreement, Source likely included the emission from a PWN. In
this case, Source would be a rotation-powered pulsar plus a PWN, or a
PWN powered by an unseen pulsar. For any conclusion, the pulsation
searches with the better time resolution are required.

There is no counterpart of Source in the ATCA radio observation (1380,
2387, 4790, and 8640 MHz \citep{dickel98}), and the flux limit was
estimated as $\sim$2 mJy in any of four frequencies. Source is quite
similar to the radio-quiet PWN around PSR B1509$-$58, which flux in
the X-ray band is factor of 2 lagrer than Source and the radio flux
limit is $\sim$1.3 mJy \citep{gaensler02,gaensler99}. Therefore, the
radio limit is consistent with our conclusion that Source is a
radio-quiet PWN.

\section{Summary}

A detailed analysis of the compact source at the center of SNR N23 in
the LMC with {\it Chandra} has been presented. N23 has $\sim 60
\arcsec \times 80 \arcsec$ diffuse emission in the $<$ 2 keV band, and an
almost point-like compact source in the $>$ 2 keV band. This source
(CXOU J050552.3$-$680141) has no counterpart at any other
wavelength. The spectrum of the source can be fitted to a power-law
function with $\Gamma$ = 2.2 $^{+0.5}_{-0.3}$, and a luminosity of 1.0
$\times$ 10$^{34}$ ergs s$^{-1}$ in the 0.5--10 keV band for a
distance of 50 kpc. Neither significant pulsation nor time variability
was detected with the time resolution of 3.2 sec observation. The
results of the fittings indicate that the source might well be a
rotation-powered pulsar and/or a pulsar wind nebula (PWN) associating
with the SNR. It implies that the progenitor of the remnant is likely
to be a massive star.

\acknowledgments
We gratefully acknowledge the {\it Chandra} team for making available
its public data used herein. We thank R. Yamazaki, K. Mori, T. Mihara,
M. Nakajima, N. Isobe and K. Makishima for their fruitful
discussions. This research made use of the SIMBAD data base, obtained
by CDS, Strasbourg, France. This work is supported in part by the
Grant-in-Aid for Young Scientists (B) of the Ministry of Education,
Culture, Sports, Science and Technology (No.17740183).

\clearpage

\clearpage
\begin{table}
\caption{Best-fit parameters for Source\tablenotemark{a}\label{tbl:fit_results}}
\begin{tabular}{lccccc}
\tableline\tableline
Power-law\tablenotemark{b} & $\Gamma$ & & ${N_{\rm H}}^{\rm LMC}$\tablenotemark{c} & Flux\tablenotemark{d} & $\chi^2$/d.o.f.\\
 & & & [$10^{20}\rm{cm}^{-2}$] & [ergs cm$^{-2}$s$^{-1}$] & \\
\tableline
with Bgd 1 & 2.2 (1.9--2.7) & &  $<$0.32 & 3.5 $\times10^{-14}$& 18.4/16\\
with Bgd 2 & 3.7 (3.0--4.8) & & 0.4 (0.1--0.9) & 7.0 $\times10^{-14}$ & 37.9/16\\
with Bgd 3 & 2.4 (1.9--2.7) & & $<$39.2 & 3.7 $\times10^{-14}$& 20.5/16\\
\tableline\tableline
Blackbody\tablenotemark{b} & $kT$ & & ${N_{\rm H}}^{\rm LMC}$\tablenotemark{c} & Flux\tablenotemark{d} & $\chi^2$/d.o.f.\\
 & [keV] & & [$10^{20}\rm{cm}^{-2}$] &  [ergs cm$^{-2}$s$^{-1}$] & \\
\tableline
with Bgd 1 & 0.26 (0.22--0.32) & & $<$0.2 & 2.0 $\times10^{-14}$& 34.7/16\\ 
with Bgd 2 & 0.20 (0.18--0.22) & & $<$0.2 & 3.3 $\times10^{-14}$& 47.7/16 \\
with Bgd 3 & 0.27 (0.24--0.29) & & $<$0.4 & 2.0 $\times10^{-14}$& 36.7/16 \\
\tableline\tableline
Plasma\tablenotemark{b} & $kT$ & Abundance & ${N_{\rm H}}^{\rm LMC}$\tablenotemark{c} & Flux\tablenotemark{d} & $\chi^2$/d.o.f.\\
 & [keV] & & [$10^{20}\rm{cm}^{-2}$] & [ergs cm$^{-2}$s$^{-1}$] & \\
\tableline
with Bgd 1 & 1.63 (1.11--4.84) & $<$1.0 & $<$0.2 & 2.6 $\times10^{-14}$& 22.2/15\\ 
with Bgd 2 & 0.65 (0.50--0.80) & $<$0.052 & $<$5.2 & 3.6 $\times10^{-14}$& 35.1/15\\
with Bgd 3 & 1.63 (0.83--2.60) & $<$0.3 & $<$0.2 & 2.7 $\times10^{-14}$ & 24.2/15\\
\tableline
\end{tabular}
\tablenotetext{a}{Values in parentheses are 90\% parameters confidence regions.}\\
\tablenotetext{b}{The assumed ${N_{\rm H}}^{\rm gal}$ was 5.6 $\times$ 10$^{20}$ cm$^{-2}$ 
\citep{dickey90}.}\\
\tablenotetext{c}{In the 0.5--10~keV band. Corrected for absorption.}\\
\tablenotetext{d}{0.3 solar abundance are used.}\\
\end{table}
\end{document}